# Direct Observation and Analysis of Low-Energy Magnons with Raman Spectroscopy in Atomically Thin NiPS$_3$

*Woongki Na,[†,#] Pyeongjae Park,[‡,§,#] Siwon Oh,[†,#] Junghyun Kim,[‡] Allen Scheie,[⊥] David Alan Tennant,[¶] Hyun Cheol Lee,[†,\*] Je-Geun Park[‡,\*] Hyeonsik Cheong,[†,\*]*

[†]Department of Physics, Sogang University, Seoul 04107, Korea

[‡]Department of Physics and Astronomy, Seoul National University, Seoul 08826, Korea

[§]Materials Science and Technology Division, Oak Ridge National Laboratory, Oak Ridge, Tennessee 37831, USA

[⊥]MPA-Q, Los Alamos National Laboratory, Los Alamos, NM 87545, USA

[¶]Department of Physics and Astronomy, University of Tennessee, Knoxville, TN 37996, USA and Department of Materials Science and Engineering, University of Tennessee, Knoxville, TN 37996, USA

[#]These authors contributed equally: Woongki Na, Pyeongjae Park, Siwon Oh

ABSTRACT

Van der Waals (vdW) magnets have rapidly emerged as a fertile playground for novel fundamental physics and exciting applications. Despite the impressive developments over the past few years, technical limitations pose a severe challenge to many other potential breakthroughs. High on the list is the lack of suitable experimental tools for studying spin dynamics on atomically thin samples. Here, Raman scattering techniques are employed to observe directly the low-lying magnon (~1 meV) even in bilayer $NiPS_3$. The unique advantage is that it offers excellent energy resolutions far better on low-energy sides than most inelastic neutron spectrometers can offer. More importantly, with appropriate theoretical analysis, the polarization dependence of the Raman scattering by those low-lying magnons also provides otherwise hidden information on the dominant spin-exchange scattering paths for different magnons. By comparing with high-resolution inelastic neutron scattering data, these low-energy Raman modes are confirmed to be indeed of magnon origin. Because of the different scattering mechanisms involved in inelastic neutron and Raman scattering, this new information is fundamental in pinning down the final spin Hamiltonian. This work demonstrates the capability of Raman spectroscopy to probe the genuine two-dimensional spin dynamics in atomically-thin vdW magnets, which can provide novel insights that are obscured in bulk spin dynamics.

**INTRODUCTION**

Two-dimensional (2D) magnetism occupies a special place in modern condensed matter physics. A complete theoretical understanding of the three fundamental Hamiltonian models—Ising, XY, and Heisenberg—has provided profound insights into a variety of magnetic phenomena and beyond.[1-4] Despite the impressive breakthroughs in the theory, progress has been much slower on the experimental side, although numerous attempts have been made from the 1960s to 1980s.[5-7] This was primarily due to the lack of suitable materials that realize the 2D magnetism, namely atomically thin sheets of magnetic materials. This gap motivated the initial exploration and eventual discovery of what are now known as van der Waals (vdW) magnets[8]. Notably, several demonstrations of monolayer magnetism have since been made in both antiferromagnetic and ferromagnetic systems within the Ising class, marking important experimental milestones.[9-11]

During the past few years, several investigations on this newly discovered category of materials have demonstrated properties distinct from three-dimensional magnetic systems,[12] reinforcing the belief that vdW magnets could usher in a new era of 2D magnetism. However, further experimental advancements are required to fully probe their unique two-dimensional characteristics undisclosed in the bulk phase. High on the wish list is the experimental technique that can measure the spin dynamics of atomically thin vdW magnets. Notably, spin dynamics provide valuable information about the spin Hamiltonian, a must for any magnetic materials. Historically, inelastic neutron scattering (INS) has played a crucial role in elucidating the spin dynamics of bulk 3D materials, greatly advancing our understanding of the spin models for bulk magnetism since the late 1950s.[13-16] However, an alternative technique is needed for atomically thin vdW samples due to a very small INS cross-section.[17]

Among several vdW magnets, NiPS$_3$ has recently gained significant interest, exhibiting XXZ-type antiferromagnetic ordering below the Néel temperature ($T_N$) of 155 K.[18, 19] Its $T_N$ decreases slightly as the number of layers is reduced from bulk to two layers but is suppressed in monolayers.[19] Remarkably, this material exhibits extremely sharp many-body magnetic excitons, as observed in photoluminescence (PL) and absorption measurements.[20, 21] Notably, these excitons are deeply connected to the local magnetic anisotropy of Ni sites,[22] which can be studied by investigating the low-energy spin dynamics of NiPS$_3$.

Optical measurements can open a new opportunity to study the spin dynamics of NiPS$_3$ samples, from bulk to few layers. Although two recent INS studies have provided an overall understanding of the spin dynamics in NiPS$_3$,[23, 24] its low-energy spin dynamics remain not fully understood. The primary challenge arises from the unclear low-energy INS profile due to pronounced instrumental resolution effects for NiPS$_3$.[24] This stems from a substantial mismatch of energy scales between the one-magnon bandwidth (55 meV) and low-energy excitations of interest (from 1 meV to 5 meV), resulting in a spectrum characterized by an excessively steep magnon dispersion which can be heavily impacted by the momentum resolution.[24] Instead, optical tools have provided clearer observations of the low-energy excitations.[25-27] For example, the temperature-dependent THz emission measurements observed a sharp peak below the Néel temperature that redshifts from ~ 5 meV at 20 K, which weakens as the temperature increases towards $T_N$.[25] On the other hand, the pump-probe polarization rotation measurements observed two low-energy excitations at 0.3 THz (~1.2 meV) and 0.9 THz (~3.7 meV) at 10 K.[26] In addition, high-magnetic field (≥7 T) electron spin resonance (ESR) measurements yielded a resonance signal at 350 GHz (1.4 meV) at 4 K and 7 T.[27] The different energy values for these reports urgently call for more accurate and reliable measurements. Moreover, since all the measurements have been conducted on bulk samples, the evolution of these low-energy magnon resonances in the atomically-thin limit, which can potentially reveal new aspects of the spin dynamics obscured in previous bulk studies, remains unanswered. For bulk magnetic materials, Raman scattering spectroscopy has been widely used to study the one- and two-magnon scattering processes, for which the work by Fleury and Loudon[28] has provided the classic principle for theoretical analysis. However, there has not yet been similar analysis on 2D magnetic systems experimentally or theoretically.

Here, we report on ultralow-frequency polarized Raman scattering measurements on atomically thin $NiPS_3$ down to monolayers. For bulk $NiPS_3$, three low-frequency modes are observed below $T_N$ at 11, 31, and 40 cm$^{-1}$ at 3.8 K, labelled $M_1$, $M_2$, and $M_3$, respectively. The three modes have different dependences on polarization and temperature. $M_1$ is relatively stronger, and so we used it to study its dependence on the thickness. By comparing with the neutron scattering data, we can confirm the origin of $M_1$ and $M_3$ as being due to the magnon excitations. On the other hand, the polarization behaviors of these modes cannot be explained based on the classic theory by Fleury and Loudon,[28] for which we developed a more general theoretical model.

**RESULTS**

**Raman Spectroscopy on Bulk $NiPS_3$.** Figure 1 shows the crystal structure of $NiPS_3$. Monolayer $NiPS_3$ has a hexagonal structure, with the Ni atoms arranged in a hexagonal lattice. For the bulk, the layers are stacked with a slight shift along the *a*-axis (zigzag direction), forming a monoclinic structure. The Ni atoms are arranged in a hexagonal lattice, each surrounded by six S atoms with a trigonal antiprismatic arrangement. The layers are held together by weak van der Waals interaction, allowing easy exfoliation. Below $T_N$, the spins are aligned in the *ab* plane, mainly along the *a*-axis, with the same-spin chains along the zigzag direction.[29]

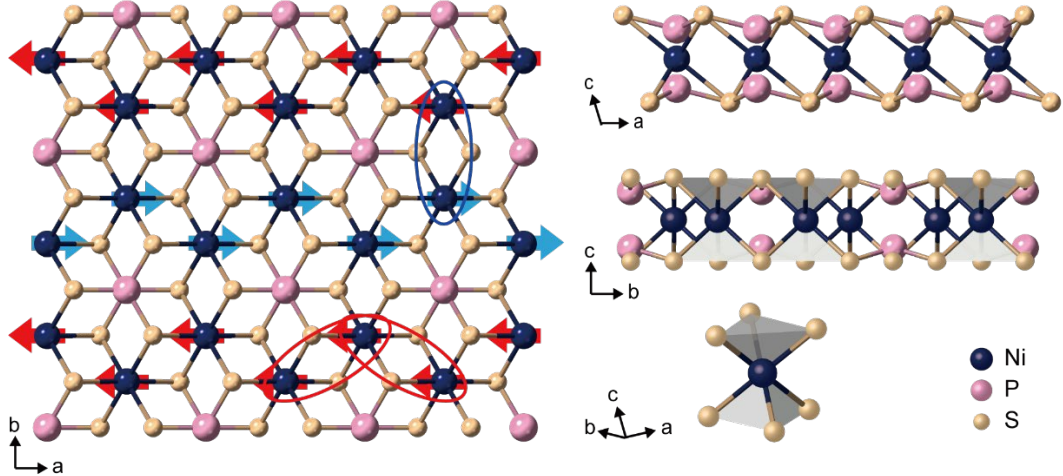

**Figure 1.** Crystal structure of NiPS$_3$. The large arrows indicate the spin directions showing the antiferromagnetic ordering. The blue and red ellipses indicate two possible types of spin pairs for the spin exchange scattering processes used in our theoretical calculations.

Figure 2a shows the temperature dependence of the polarized Raman spectra of a ~170-nm-thick bulk NiPS$_3$ sample measured in $z(xx)\bar{z}$ and $z(xy)\bar{z}$ configurations, where z and $\bar{z}$ indicate the directions of incident and scattered photons, respectively, and the variables inside the parenthesis indicate the polarization directions of the incident and scattered photons. Here, we set the *x*-direction 45° relative to the *a*-axis of the crystal determined by the polarization dependence of the photoluminescence at ~1.48 eV with those of the split Raman peaks at ~180 cm$^{-1}$ (Figure S1),[20] the splitting of which ($\Delta P_2$) is a leading indicator of the antiferromagnetic ordering.[19] At 290 K, the low-frequency Raman spectrum for the parallel [$z(xx)\bar{z}$] polarization configuration is dominated by the quasi-elastic scattering (QES) signal due to spin fluctuations, which is suppressed at low temperatures.[19] As the temperature is lowered below $T_N$ of ~155 K, a sharp peak ($M_1$) and two weaker peaks ($M_2$ and $M_3$) emerge. These peaks become stronger and blueshift as the temperature is further lowered. Although $M_1$ overlaps with the QES signal, it can be resolved for temperatures of ~140 K and below. At 3.8 K, $M_1$ is at 11 cm$^{-1}$, corresponding to 0.33 THz or 1.4 meV, $M_2$ at 31 cm$^{-1}$ (0.93 THz or 3.8 meV), and $M_3$ at 42 cm$^{-1}$ (1.3 THz or 5.2 meV). The frequency of $M_1$ is also close to what was observed in the pump-probe polarization rotation measurements[26] or the ESR measurements,[27] and $M_2$ was also observed in the pump-probe measurements.[26] At the same time, $M_3$ is similar to that observed in the THz emission experiments.[25]

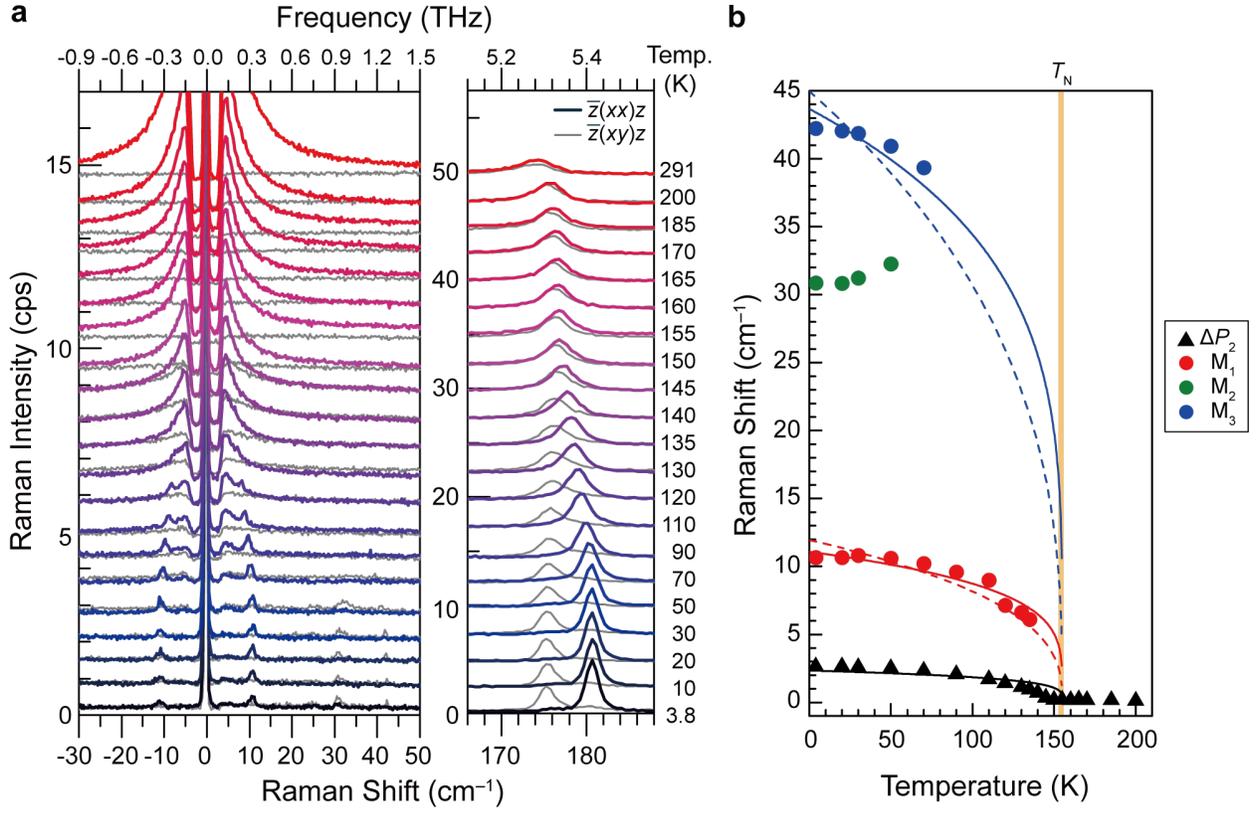
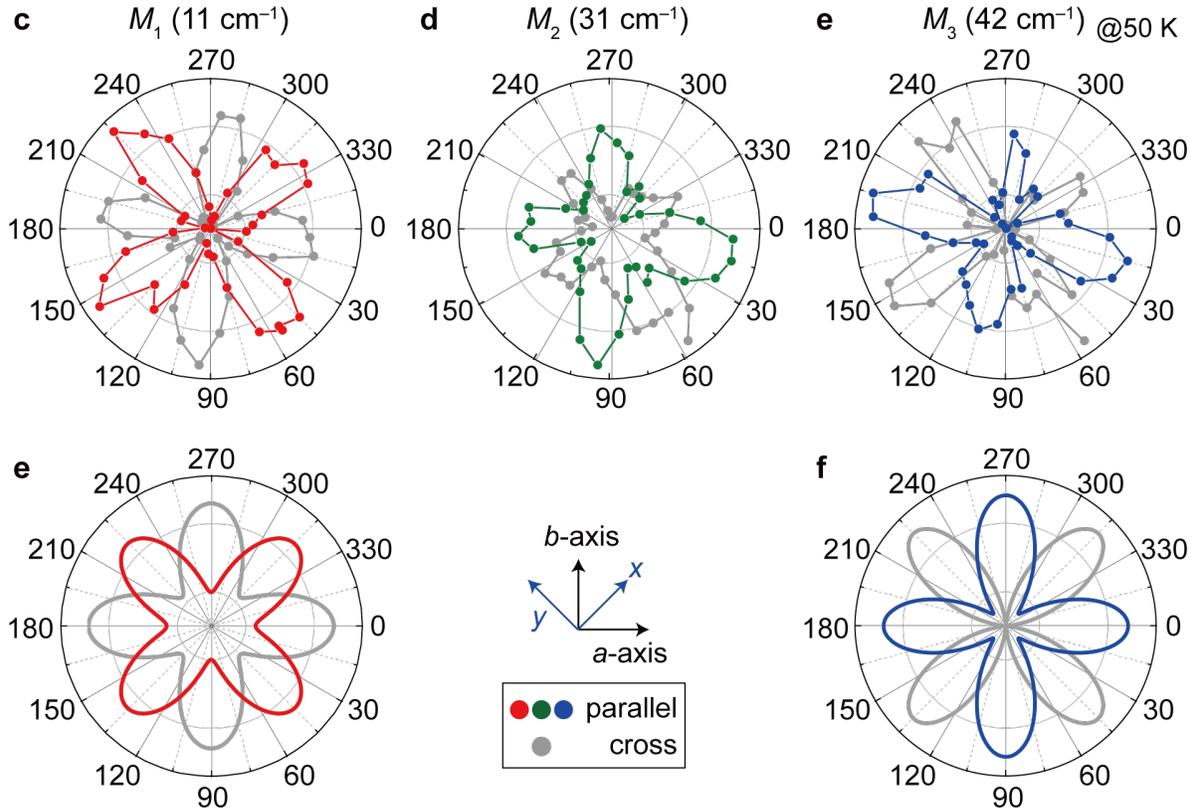

**Figure 2.** Low-frequency Raman spectra of bulk NiPS$_3$. a) Temperature dependence of the polarized Raman spectra of bulk (~170 nm) NiPS$_3$. The colored spectra are measured in $z(xx)\bar{z}$ configuration and the grey spectra in $z(xy)\bar{z}$. At low temperatures, M$_1$ is observed in $z(xx)\bar{z}$, whereas M$_2$ and M$_3$ are observed in $z(xy)\bar{z}$. The splitting (ΔP$_2$) of the mode at ~180 cm$^{-1}$ for the two polarizations indicates antiferromagnetic ordering.[19] b) Temperature dependence of the M$_1$, M$_2$, M$_3$, and ΔP$_2$. The extrapolated fitting curves for XY (solid) and Heisenberg (dashed) models are shown. c,d,e) Polarization dependence of the intensities of M$_1$, M$_2$, and M$_3$ in the parallel (red, green, blue) and the cross (grey) polarization configurations, respectively. e) Calculated polarization dependence for the red spin-exchange paths in Figure 1. The red (grey) trace is for the parallel (cross) polarization configuration. f) Calculated polarization dependence for the blue spin-exchange paths in Figure 1. The blue (grey) trace is for the parallel (cross) polarization configuration.

Since M$_2$ and M$_3$ are too weak in few-layer samples, we mainly analyzed M$_1$ for the few-layer cases. This peak's excitation-laser-wavelength dependence was surveyed (Figure S2), and it appears strongest for the excitation wavelength of 632.8 nm (1.96 eV), probably due to the resonance with the optical gap of NiPS$_3$ at 1.8 eV.[30] For all the data presented below, we used the 632.8-nm excitation. Figure 2b shows the temperature dependence of M$_1$, M$_2$, M$_3$, and ΔP$_2$. The solid curves show the temperature dependence of $[1-T/T_N]^\beta$ with the exponent $\beta$ of 0.23 for the XY model,[31] whereas the dashed curves for the 3-dimensional Heisenberg model[32] with $\beta$ of 0.369 do not show similar agreement. The close agreement with the XY model ascertains that M$_1$ and M$_3$ correlate with the antiferromagnetic ordering. Recent magneto-optical measurements on bulk NiPS$_3$ also indicated that M$_1$ and M$_3$ are of magnetic origin.[33] On the other hand, the frequency of M$_2$ increases with temperature, which was also observed in the previous pump-probe data.[26] This indicates that M$_2$ is not of the same origin as M$_1$ and M$_3$. We note that there is a suggestion that M$_2$ is due to a zone-folded phonon.[34]

Figure 2c shows the polarization dependence of $M_1$ measured in the parallel- (red) and cross- (grey) polarization configurations. In parallel (cross) polarization, the analyzer for the scattered photon is set in the direction parallel (cross) to the polarization of the incident photons, and the relative angle between the incident polarization and the crystal axes was rotated. Here, 0° corresponds to the *a*-axis direction of the crystal. The polarization behaviors of $M_2$ or $M_3$ are opposite to that of $M_1$ (Figure 2d,e), exhibiting correlations with the underlying crystal symmetry. Given the same magnon nature of both $M_1$ and $M_3$, this completely different polarization dependence is most unexpected. To say the least, this observation contradicts the classic theory on light scattering by magnons by Fleury and Loudon,[28] which predicts that the light scattering by one magnon is purely antisymmetric, and the one-magnon Raman scattering vanishes in parallel polarization, regardless of the directions of the incident and scattered wave vectors.[28] Their predictions are incompatible with our experimental data, which strongly depend on the crystal structure and are clearly not antisymmetric. Therefore, we are led to develop our own theoretical model to explain the experimental results.

**Theoretical Analysis on Raman Selection Rule.** Shastry and Shraiman[35] studied the Raman scattering of Mott-Hubbard systems based on the tight-binding one-band Hubbard model. In their approach, the basic mechanism for the magnetic Raman scattering is the photon-induced spin super-exchange, and the electron-photon interaction is described by the electron hopping Hamiltonian multiplied by the Peierls phase: (spin indices suppressed below)

$$H_{\text{hop-em}} = c^{\dagger}_{\mathbf{R}_i \alpha} t^{\alpha\gamma}_{ij} \exp\left(-i\frac{e}{\hbar c}\int_{\mathbf{R}_j}^{\mathbf{R}_i} \mathbf{A}(\mathbf{r}) \cdot d\mathbf{r}\right) c_{\mathbf{R}_j \gamma}, \tag{1}$$

where $\mathbf{R}_i$ indicates the position of the *i*-th lattice site, and $\alpha, \gamma$ denote general quantum numbers such as orbitals. $c^{\dagger}_{\mathbf{R}_i \alpha}$ is the electron creation operator at the site $\mathbf{R}_i$ with quantum number $\alpha$. $t^{\alpha\gamma}_{ij}$ is the hopping amplitude of the tight-binding model, and $\mathbf{A}(\mathbf{r})$ is the vector potential of the photon. The photon polarization is directly coupled to the crystal structure in this formulation. They then derived an effective Hamiltonian for Raman scattering by magnetic excitation by treating the electron-photon coupling perturbatively in the second order within the framework of strong coupling expansion of super-exchange. Their result, Equation 8 of Ref. [35], (a missing term is inserted below) is given by

$$H_{\text{eff}} = \sum_{\mathbf{R}} (\frac{1}{4} - \mathbf{S}_{\mathbf{R}} \cdot \mathbf{S}_{\mathbf{R}+\mu}) \left( \frac{t^2}{U - \hbar\omega_{\text{incoming}}} + \frac{t^2}{U + \hbar\omega_{\text{scattered}}} \right) [\mathbf{A}^{\text{scattered}} \cdot \boldsymbol{\mu}][\mathbf{A}^{\text{incoming}} \cdot \boldsymbol{\mu}], \quad (2)$$

where $\boldsymbol{\mu}$ is a vector connecting neighbors and $\mathbf{S}_{\mathbf{R}}$ is a spin-1/2 operator at site $\mathbf{R}$. $U$ is the on-site Coulomb repulsion, and *t* is the hopping parameter of the one-band Hubbard model, respectively. $\omega_{\text{incoming (scattered)}}$ is the frequency of the incoming (scattered) phonon. It is to be noted that, unlike the classic theory of Fleury and Loudon, the coupling between polarization and crystal structure is now explicit in this new derivation. Clearly, the scattering is no longer *antisymmetric* under the interchange of the polarizations of incoming and scattered photons. However, their result, Eq. (2) based on one-band Hubbard model, cannot yet be directly applied to our case of NiPS$_3$, mainly due to the differences in the valency of Cu and Ni and the crystal structure.

Here, we recall the previous theoretical work by Koshibae, Ohta, and Maekawa on NiO$_2$.[36] Their tight-binding Hamiltonian can be adapted to NiPS$_3$ by modifying the underlying crystal structure. Our goal is to derive the magnetic Raman scattering Hamiltonian based on the adapted

Hamiltonian in the framework of Shastry and Shraiman's approach. Considering the complexity of NiPS$_3$, we seek a tractable minimal model which can describe some features of magnetic excitations. We presume the essential elements to be (1) hopping between Ni and S ions, (2) on-site Coulomb repulsion of Ni ions, and (3) spin-orbit interaction, which is needed for the single-ion magnetic anisotropy responsible for the planar spin configuration.[37] We ignore (1) trigonal distortion, (2) Coulomb repulsion and spin-orbit interaction of S ions, (3) direct hopping between Ni ions and between S ions, and (4) all inter-site Coulomb repulsions. It turns out that the qualitative feature of non-antisymmetric scattering polarization dependence can be understood by taking the Ni-S hopping and Ni on-site Coulomb repulsion on the qualitative consistency level. Unfortunately, the inclusion of other elements, in particular the spin-orbit interaction and the trigonal distortion,[38-41] render the analytic calculations almost impossible. Evidently the above-ignored elements can play important roles in a more complete description of magnetic excitations.

The basic structure of the effective magnon scattering Hamiltonian obtained from the tight-binding model adapted for NiPS$_3$ is

$$\hat{H}_{basic} = P_0 \hat{H}_{e\text{-photon}} \hat{G} \hat{H}_{hop} \hat{G} \hat{H}_{hop} \hat{G} \hat{H}_{e\text{-photon}} P_0, \tag{3}$$

where $P_0$ is the projector onto the subspace of empty S ion holes and Ni spin-triplet states with the energy $E_0$, and $\hat{G}$ is the resolvent for the complementary space of $P_0$ (see Methods). The spins can be exchanged in *two distinct mechanisms: via doubly occupied Ni ion or via doubly occupied S ion*. On top of this, the edge-sharing crystal structure allows several spin exchange paths, depending on the mechanism. The scattering polarization dependence can be extracted from Eq. (3) by reorganizing electron operators in the form of a spin operator (with spin 1). (We mention that the spin-orbit interaction is not considered for the computation of the effective Hamiltonian.)

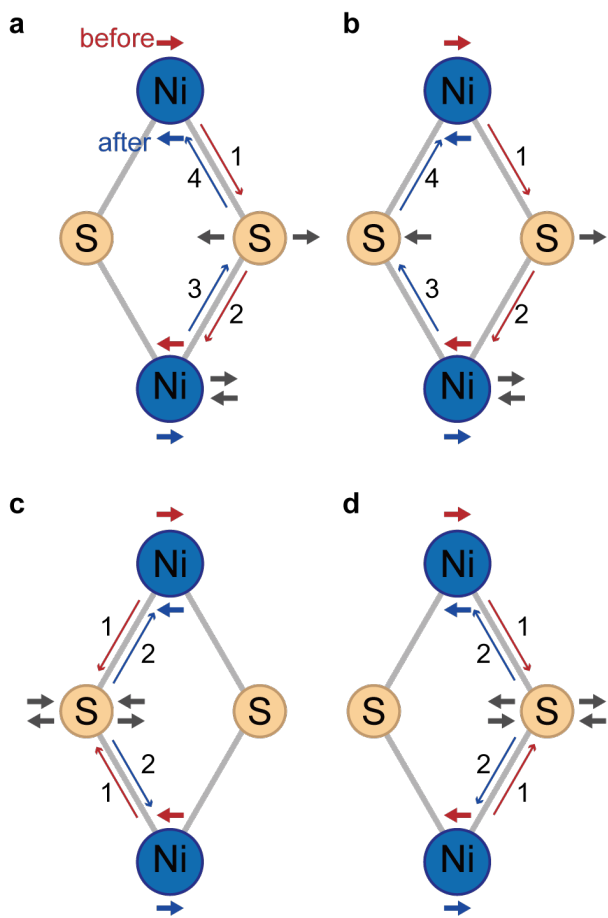

**Figure 3.** Schematics of typical spin-exchange in NiPS$_3$ for the opposite-spin pair indicated by the blue ellipse in Figure 1. Spin is exchanged via the doubly occupied Ni ion (a, b) or the doubly occupied S ion (c,d), respectively. The thick red and blue arrows indicate the electron spin before and after the spin-exchange, respectively. The thick gray arrows indicate the spins of the intermediate states. The thin red and blue arrows between Ni and S ions indicate the forward and return paths, respectively. For spins at a doubly occupied Ni ion, there are two possible return paths: either hopping to the right (a) or the left (b) S ion.

Figure 3 illustrates the spin exchange mechanism for the opposite-spin pair (marked by blue ellipse in Figure 1). Spin is exchanged via doubly occupied Ni or S ions. There are four possible exchange paths (and their equivalents). For example, in Figure 3a, the electron with right-spin in the upper Ni ion is scattered to the nearby Ni ion with left-spin. From this doubly occupied Ni ion, an electron with left-spin is returned to the upper Ni ion, completing the spin exchange. For the parallel-spin pair (marked by red ellipses in Figure 1), the overall exchange mechanism is similar with the consideration of multiple orbitals in the doubly occupied Ni or S ions. Note that Figure 3 depicts spin-1/2 exchanges, whereas the spin-1 scattering Hamiltonian (Eq. (9) in Materials and Methods) is obtained by considering $e_g$ multiple orbitals and the Wigner-Eckart theorem (see Supporting Information Eq. (S13) and accompanying explanation).

By combining the contributions for a ferromagnetic Ni ion pair, we can obtain a polarization dependence consistent with $M_1$ (Figure 2e). In contrast, those from an antiferromagnetic pair are consistent with $M_3$ (Figure 2f). This comparison shows that the spin exchange scattering paths in the zigzag direction (same spins) dominate the $M_1$ magnon. Those in the armchair (opposite spins) directions are more critical for the $M_3$ magnon. We stress that this observation is an entirely new piece of information. Usually, the super-exchange generates antiferromagnetic coupling between spins. However, from the viewpoint of the effective scattering Hamiltonian, the sign of the *effective* exchange constant can be both positive and negative, as can be seen in Eq. (2). For more rigorous calculations, one has to consider the complicated sign patterns of the hopping amplitude between *d* orbital of Ni ion and *p* orbital of S ion in trigonally distorted octahedron geometry and the spin-orbit interaction which can be responsible for the zigzag planar spin configuration. This is well beyond the scope of the current work.

**Raman Spectroscopy on Few-layer NiPS$_3$.** Figure 4a shows the dependence of $M_1$ at 50 K on the number of layers along with that of a bulk sample. The polarization was chosen so that the intensity of $M_1$ is at its maximum. The full set of spectra is shown in Figure S3 and S4. The evolution of the interlayer breathing mode (■) with the layer number is consistent with the previous report.[19] $M_1$ is observed for thicknesses down to 2L, whilst it is absent from the spectrum of 1L. This is also consistent with the suppression of the magnetic ordering[19] in 1L NiPS$_3$ and serves as additional evidence that this peak is indeed a result of the antiferromagnetic ordering.

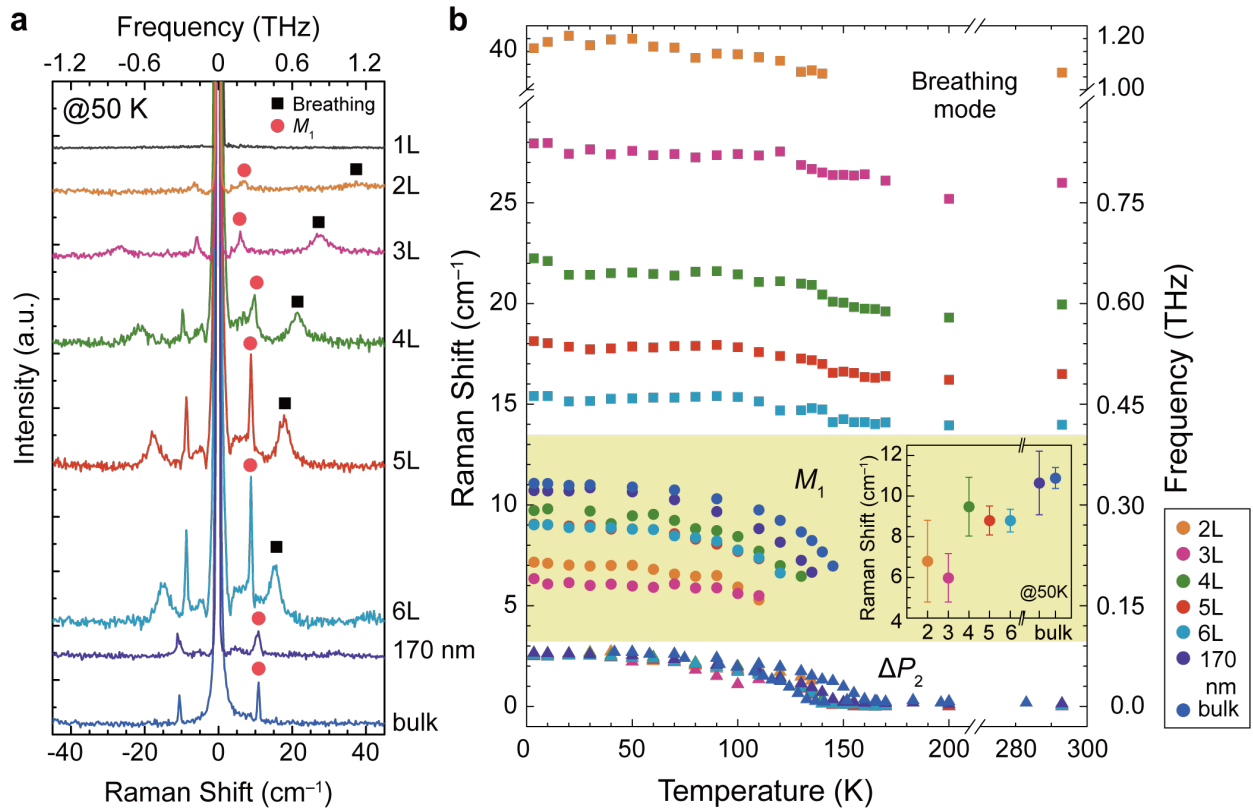

**Figure 4.** Temperature dependence of low-frequency Raman modes of few-layer NiPS$_3$. a) Layer-number dependence of low-temperature Raman spectra. The $M_1$ and interlayer breathing modes are marked with red circles and black square symbols, respectively. b) Temperature dependence of the Raman modes in the low-frequency range for different numbers of layers. $\Delta P_2$ is also plotted for each number of layers for comparison. The inset shows the thickness dependence of the $M_1$ magnon mode.

Unfortunately, $M_2$ and $M_3$ are too weak to be resolved reliably in few-layer samples, so we cannot examine their thick dependence in detail. But $M_1$ shows a clear dependence on the number of layers: the peak redshifts as the thickness decreases, although the redshift does not scale linearly with the thickness. Potential origins for the decrease in the peak frequency are discussed below. The temperature dependences of $M_1$ for few-layer samples are similar to that for the bulk sample and can be described with the same exponent of 0.23 (Figure 4b and Figure S5a). For thinner samples, $M_1$ becomes unresolved as the temperature increases toward $T_N$ because of the enhanced QES signal near the magnetic transition temperature[19] (see Figure S4). The intensity of $M_1$ tends to increase with temperature up to ~40 K and then decrease again when the temperature is further raised beyond ~80 K (Figure S5b). This temperature dependence is not understood yet. Interestingly, the breathing modes also blueshift below $T_N$. This is because the additional magnetic interaction between the layers below $T_N$ increases the 'spring constant' between the layers.

**Inelastic Neutron Scattering Analysis.** We complement our analysis with the INS measurement results to better understand the low-energy magnetic excitations. Previous INS studies successfully captured the overall magnon-spectra of $NiPS_3$ within the [$H$, $K$, 0] plane (in reciprocal lattice units, r.l.u.), including the gapped low-energy magnon mode at approximately 5 meV (e.g., see Q = [0, 1, 0] in Figure 5a, where Q denotes a momentum).[23, 24] Notably, this energy corresponds to the excitation energy of the $M_3$ mode. However, the spectra in the [$H$, $K$, 0] plane failed to identify a signal below 5 meV, i.e., the $M_1$ mode. After having obtained the Raman data shown earlier, we examined our data more carefully and observed a clear magnon dispersion down to approximately 1.5 meV at the magnon zone centers on the [$H$, $K$, 1] plane.

Figure 5b shows such observations at Q = [0, 1, 1], and a similar spectrum at Q = [1/2, 1/2, 1] – an equivalent Q position to [0, 1, 1] but simply corresponding to a different magnetic domain – is shown in Supporting Information (Figure S6). These results suggest that $M_1$ (1.5 meV or 12 cm$^{-1}$) and $M_3$ (5 meV or 40 cm$^{-1}$) are the single-magnon excitations at the $\Gamma$ point. On the other hand, there is no clear indication of a magnetic excitation corresponding to the energy of $M_2$ (3.8 meV). Thus, the $M_2$ mode is not likely to be a single-magnon excitation. This is consistent with a conventional dipole spin-wave theory, which yields only two magnon eigenvalues at the $\Gamma$ point as two spin sublattices are present in NiPS$_3$.

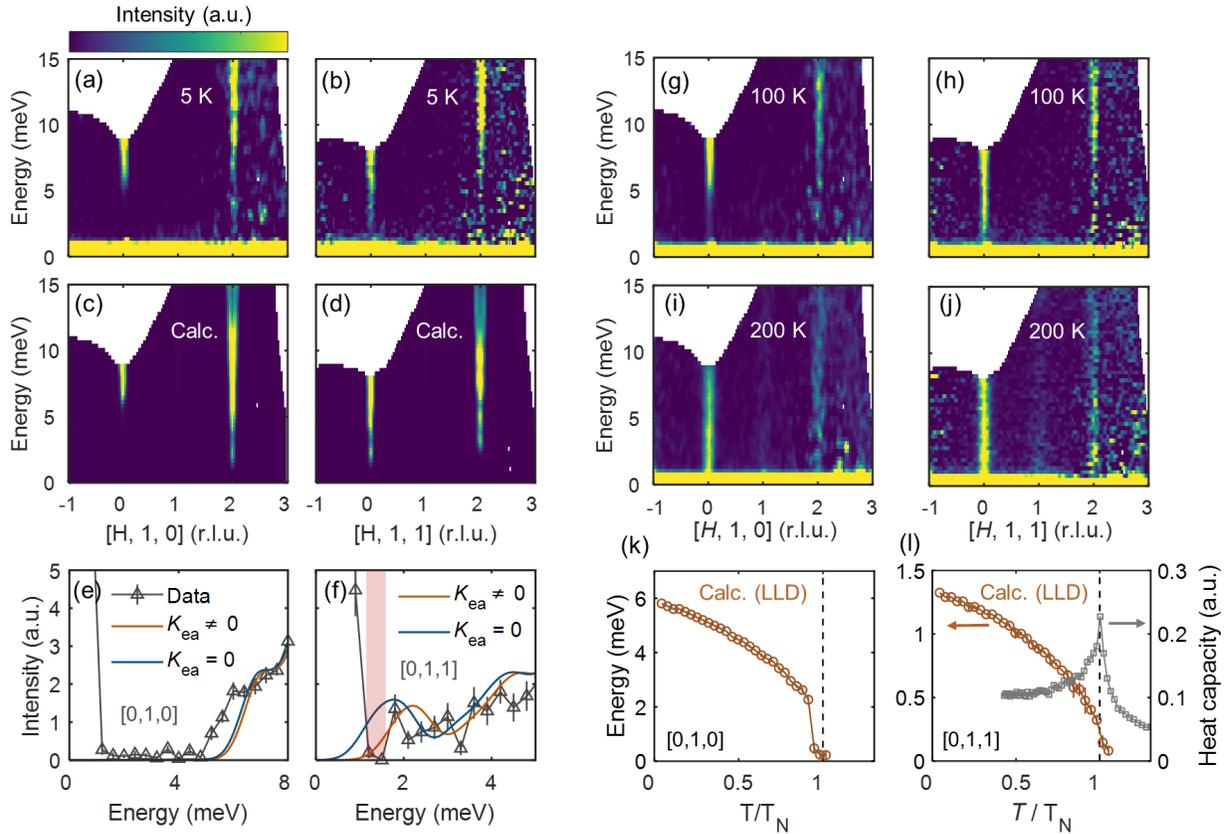

**Figure 5.** Low-energy magnon spectra of NiPS$_3$. a,b) The magnon spectra measured by inelastic neutron scattering (INS) at 5 K. c,d) The magnon spectra calculated by linear spin-wave theory (LSWT, see Methods), each corresponding to (a) and (b), respectively. e,f) The constant-**Q** cuts of the INS data in (a) and (b) at **Q** = [0, 1, 0] and [0, 1, 1] (r.l.u.), each demonstrating gapped magnon

modes at ~5 meV and ~1.5 meV. Orange and blue solids curves are the LSWT calculation results with and without easy-axis anisotropy $K_{ea}$, respectively. The simulation results in (c)-(f) include instrumental resolution and experimental binning width effects. g,h,i,j) The magnon spectra measured at 100 K (< $T_N$) and 200 K (>$T_N$). k,l) The temperature dependence of the magnon energy gap at **Q** = [0, 1, 0] and **Q** = [0, 1, 1] calculated by Landau-Lifshitz dynamics (LLD) simulation (see Methods). Grey data points are the heat capacity calculated from the same simulation, showing the transition temperature of the theoretical NiPS$_3$ spin system.

For a better understanding on the two gapped modes assigned as M$_1$ and M$_3$, we calculated the theoretical magnon spectra of NiPS$_3$ using the linear spin-wave theory (LSWT, see Methods). The spin Hamiltonian used in our calculations was based on previous studies:[23, 24]

$$H = J_n \sum_{(i,j)_n} \mathbf{S}_i \cdot \mathbf{S}_j + K_{ep} \sum_i (S_i^z)^2 + K_{ea} \sum_i (S_i^x)^2 \qquad (4)$$

where $J_n$ and $(i,j)_n$ denote the exchange constant and the site indices for the $n^{th}$ nearest neighbors, and $K_{ep}$ (> 0) and $K_{ea}$ (< 0) are the magnitudes of easy-plane (the *a-b* plane) and easy-axis (*x // a*-axis) anisotropy, respectively. We adopted the same exchange and single-ion anisotropy parameters as those in Ref. 24. As shown in Figure S7, Supporting Information, this model successfully yields magnon eigenvalues of 1.28 meV (= 10.32 cm$^{-1}$) and 5.97 meV (= 48.15 cm$^{-1}$) at the Γ point, each aligning with the energy of M$_1$ and M$_3$. Notably, each eigenmode originates from non-zero $K_{ea}$ and $K_{ep}$ in Eq. (4), respectively, indicating that the formation of M$_1$ (M$_3$) is associated with the easy-axis (easy-plane) anisotropy at the Ni sites.

Our dynamical structure factor calculations further reveal that only the 5.97 meV mode is visible at **Q** = [0, 1, 0] (Figure 5c or Figure S7), consistent with our observation (Figure 5a). On the other hand, both modes possess sizable structure factor and thus should appear together at **Q** =

[2, 1, 0] or [0, 1, 1] (Figure S7). However, such a double-gap structure anticipated from our calculation does not manifest clearly in the INS spectra, as depicted in Figure 5a and b. The reason behind this discrepancy is the effect of momentum resolution and experimental bin width that are particularly significant in the low-energy excitation spectra of NiPS$_3$, as described in the Introduction or Ref. [24]. Figure 5c,d show the calculated magnon spectra with such resolution effects taken into account, indeed bearing strong similarities to the measured INS spectra (Figure 5a,b).

The presence of M$_1$ in the INS data is more evident in its constant-**Q** cut at **Q** = [0, 1, 1] (Figure 5f). While a quasi-elastic background signal dominates the spectrum for $E < 1$ meV, the data points within a red-shaded region in Figure 5f indicate an energy gap of approximately 1.5 meV (= 12 cm$^{-1}$). This gapped feature was further verified by comparing it with the calculation from $K_{ea} = -0.01$ meV and $K_{ea} = 0$, each representing a gapped and gapless magnon model, respectively. Both the gapped (solid orange curve in Figure 5f) and gapless (solid blue curve in Figure 5f) magnon models result in a gapped feature due to the significant resolution effect mentioned earlier. However, the gapless magnon model fails to explain the vanishing intensity near 1.5 meV (the red-shaded region). This again supports our interpretation that M$_1$ corresponds to a gapped single-magnon excitation stemming from $K_{ea}$.

Temperature dependence of these low-energy magnetic excitations, examined through INS and theoretical calculations, not only corroborates the single-magnon nature of M$_1$ and M$_3$ but also reveals a non-trivial character beyond semi-classical spin dynamics theory. Figure 5g-j show the measured INS spectra at 100 and 200 K. Indeed, increased thermal fluctuations cause the M$_1$ and M$_3$ magnon modes to get subdued significantly at higher temperatures (Figure 5g,h) and eventually close their energy gap for $T > T_N$ (Figure 5i,j), similar to the Raman spectroscopy result (Figure 2). The corresponding spin dynamics simulations at finite temperatures were conducted using semi-classical Landau-Lifshitz dynamics (LLD) (see Methods). Similar to the LSWT calculation, we employed the spin Hamiltonian described in Equation 4 for this simulation. The results shown in Figure 5k and l present that the energy gaps behave like an order parameter of the phase transition at $T_N$. However, the LLD simulation indicates a more gradual increase in magnon energy below $T_N$ (Figure 5l) compared to the observed behavior of $M_1$. Hence, while our spin model adequately describes the presence of the $M_1$ magnon mode and its qualitative dissipation by thermal fluctuations, $M_1$ possesses additional characteristics not captured by the semi-classical spin dynamics theory. Importantly, this aligns with the intensity anomaly of the $M_1$ peak observed on the [$H$, $K$, 0] plane, which cannot be modelled accurately by LSWT.[24]

The observed thickness dependence of $M_1$ warrants further discussion, particularly with regards to its relevance to the easy-axis anisotropy ($K_{ea}$) in Eq. (4). Phenomenologically, the sudden decrease in $M_1$'s energy in bi- and tri-layer NiPS$_3$ can be interpreted as a reduction in either $|K_{ea}|$ or the spin expectation value <$S$> (i.e., the magnitude of the ordered moment). While a change in <$S$> would arise to some extent due to the enhanced fluctuations in reduced dimensions, the change in $|K_{ea}|$ is also expected to take place in the nearly 2D limit, given its origin inferred from symmetry arguments. While each layer in NiPS$_3$ conforms to trigonal symmetry (e.g. a honeycomb lattice), the monoclinic stacking in the bulk structure lowers the entire symmetry of the system to two-fold. Thus, $K_{ea}$, which is not compatible with the trigonal geometry of a single isolated layer, is presumably induced by the symmetry lowering by the monoclinic inter-layer environment. However, in bi- and tri-layer NiPS$_3$, a significant portion of the sample (e.g. 33 % in bi-layer nanoflakes) interfaces with vacuum or substrate instead of another NiPS$_3$ layer. Thus, the influence of monoclinic stacking is expected to decrease a lot, leading to a sizable change in $|K_{ea}|$. Notably, such modifications driven by reduced dimensions can only be elucidated through spin dynamics measurements of atomically-thin flakes – an analysis beyond the reach of an INS technique. On the other hand, the evolution of $M_3$ with decreasing thickness is anticipated to be less dramatic, as the easy-plane anisotropy $K_{ep}$, the source of $M_3$, is compatible with the trigonal symmetry inherent to each layer. Finally, unlike $M_1$ and $M_3$ we failed to observe any feature in our inelastic neutron scattering data that can be associated with the $M_2$ peak in the Raman data: which reinforces that it is more likely to be of phonon origin.

**CONCLUSION**

NiPS$_3$ is arguably the most exciting vdW antiferromagnet with several noticeable features, including the highly narrow and linear polarized exciton peak. It is believed that in NiPS$_3$, all three degrees of freedom: spin, charge, and lattice, are entangled with one another. Using the high-resolution ultralow-frequency Raman technique, we succeeded in measuring low-lying magnons down to bilayer NiPS$_3$, which has so far remained in the realm of the unachievable despite its fundamental importance. By comparing with inelastic neutron scattering data, we established the magnon nature of these low-energy modes beyond doubt. The Raman signals due to these magnons exhibit striking polarization dependences that according to our new theoretical model, are intricately related to the spin-exchange scattering paths of these magnons. Our work opened the door to a new way of studying spin dynamics down to atomically thin layers of vdW magnets, which can be applied to many vdW and other 2D magnets such as FePS$_3$ which also shows magnon signals in the Raman spectrum.[42]

**MATERIALS AND METHODS**

**Sample Preparation and Characterization.** NiPS$_3$ crystals were grown by the chemical vapor transport (CVT) method, as explained in Ref. [43] and Ref. [24]. Inside an argon-filled glove box, elemental powders (purchased from Sigma-Aldrich) of nickel, phosphorus, and sulfur were weighed and mixed in a stoichiometric ratio. Few-layer NiPS$_3$ samples were mechanically exfoliated onto 285-nm SiO$_2$/Si substrates using scotch tape from single-crystal flakes. The thickness of each sample was determined by the optical contrast of microscope images and Raman spectroscopy. The Néel vector of the antiferromagnetically ordered phase was determined from the polarization dependence of the magnetic exciton photoluminescence signal at 1.48 eV.[44]

**Raman Spectroscopy.** The temperature-dependent low-frequency Raman spectra of bulk and few-layer NiPS$_3$ were measured in a micro-measurement system using a closed-cycle He cryostat (Montana instruments). The laser beam was focused to a focus of ~1 μm in diameter by a 40× microscope objective lens (0.6 N.A.) with a power below 100 μW to avoid damage to the sample, and the scattered light was collected and collimated by the same objective. The excitation light was the 632.8-nm line of a He-Ne laser except for the comparison of the excitation wavelengths in Figure S1. To access the low-frequency range below 100 cm$^{-1}$, volume holographic filters (OptiGrate) were used to clean the laser lines and to reject the Rayleigh-scattered light. For parallel (cross) polarization, the analyzer angle was set such that photons with polarization parallel (cross) to the incident polarization pass through. Another achromatic half-wave plate was placed in front of the spectrometer to keep the polarization direction of the signal entering the spectrometer constant with respect to the groove direction of the grating. The Raman signal was dispersed by a Jobin-Yvon Horiba iHR550 spectrometer (2,400 grooves/mm) and detected with a liquid-nitrogen-cooled back-illuminated charge-coupled-device (CCD) detector.

**Single-Crystal INS Measurements.** We conducted the inelastic neutron scattering (INS) experiment using the SEQUOIA time-of-flight spectrometer[45, 46] at the Spallation Neutron Source (Oak Ridge National Laboratory, USA).[47] For the experiment, we co-aligned 26 single-crystal pieces (~ 2.41 g) of NiPS$_3$ on three circular aluminum plates (see Ref. [24]). We collected the data at 5, 100, and 200 K with three incident energies: 28, 60, and 100 meV. The sample was mounted in the geometry of the (H K 0) scattering plane and was rotated during the measurement to collect the spin waves over the full Brillouin zone. A background signal was estimated by measuring an empty sample holder's signal and was subtracted from our INS data. We used Horace software for the data analysis.[48] The data were symmetrized based on the in-plane mirror operation: that is, [H, K, L] → [H, -K, L]. The data integration range of the plots in Figure 5a and b and Figure 5g and h is $\pm 0.05$ r.l.u. for [0, K, 0] direction and $\pm 0.15$ r.l.u. for the direction perpendicular to the [H, K, 0] plane. The integration range of the constant-**Q** cuts in Figure 5e and f is $\pm 0.04$ r.l.u. for [H, 0, 0] and [0, K, 0] directions and $\pm 0.15$ r.l.u for the direction perpendicular to the [H, K, 0] plane.

## THEORETICAL CALCULATIONS

**Spin-Wave Calculations.** We calculated the magnon eigenvalues and INS cross-sections of NiPS$_3$ using linear spin-wave theory. For this calculation, we used the SpinW library.[49] For a precise comparison with the measured INS spectra, we applied the instrumental resolution and the effect of experimental bin width to our simulation. The calculation results are shown in Figure 5c-f and Figure S7. We simulated the spin system based on Landau-Lifshitz dynamics (LLD) to calculate the low-energy magnon modes at finite temperatures. This was done using the calculation package Su(n)ny, whose details can be found in Refs. [50, 51, 52]. For bulk NiPS$_3$, we prepared a 100×100×2 supercell (80,000 Ni$^{2+}$ sites) and sampled its time evolution governed by LLD. To calculate the magnon mode at ~1.28 meV (5.97 meV), a Langevin time step and a damping constant were set to d$t$ = 0.2/$J_3$ = 0.0143 meV$^{-1}$ (d$t$ = 0.1/$J_3$ = 0.0072 meV$^{-1}$) and $\lambda = 0.1$, respectively. The sampling was begun after evolving the system over 4,000~8,000 Langevin time steps for equilibration, which depends on the length of a single Langevin time step. Based on the sampled results, we calculated the dynamical structure factor of the spin system and analyzed the energy of a magnon peak at **Q** = [0, 1, 0] and [0,1,1] (r.l.u.). We repeated this calculation 10 times at each temperature and used the averaged values. The result is shown in Figure 5k and l.

**Calculation of Polarization Dependence.** First, let us describe the tight-binding model Hamiltonian for NiPS$_3$. Adopting the hole picture for the electronic states of Ni and S ions, the vacuum is the filled 3d levels of Ni and 2p levels of S. Then Ni ions will have two holes in $e_g$ orbitals ($x^2 - y^2$, $3z^2 - r^2$, denoted by 0, 1) in the spin-triplet state. Hund's rule on-site Coulomb interaction is implicitly assumed to enforce the spin-triplet ground state at each Ni ion (this is implemented by Wigner-Eckart theorem, see chapter 7 of Ref. [53] and Ref. [36]). The tight-binding Hamiltonian consists of on-site terms and hopping terms. Recall that the photon couples to an electron through the hopping terms.

$$\hat{H} = \hat{H}_{\text{on-site}} + \hat{H}_{\text{hop}} + \hat{H}_{\text{e-photon}}. \tag{5}$$

The on-site Hamiltonian consists of the site energy of Ni and S ions, Hubbard U repulsion, inter-orbital repulsion, and spin-orbit interaction. The detailed expressions are in Supplementary Information. All parameters of these Hamiltonians should be regarded to be phenomenological. The on-site spin-orbit interaction is *not* going to be included in the computation of effective scattering Hamiltonian for the sake of simplicity. *This is a grave approximation* and can be potentially problematic.

Next, take the crystal structure into account. Ni ion sites form a honeycomb lattice. Also, each Ni ion sits at the center of a hexagon formed by S ions. This crystal structure is edge-sharing so that each S ion is shared by two neighboring Ni ions; hence, it allows *multiple spin super-exchange paths*. To derive the effective Hamiltonian, we apply the strong coupling expansion of super-exchange in the second order of $\hat{H}_{\text{e-photon}}$ (the second order light scattering) and the second order of $\hat{H}_{\text{hop}}$ (for the intermediate states created by photon involving the doubly occupied states). The basic structure of the effective magnon scattering Hamiltonian is

$$\hat{H}_{\text{basic}} = P_0 \hat{H}_{\text{e-photon}} \hat{G} \hat{H}_{\text{hop}} \hat{G} \hat{H}_{\text{hop}} \hat{G} \hat{H}_{\text{e-photon}} P_0, \tag{6}$$

where $P_0$ is the projector onto the subspace of empty S ion holes and Ni spin-triplet states with the energy $E_0$.

$$\hat{G} = \frac{\text{Id} - P_{E_0}}{E_0 - \hat{H}_{\text{on-site}}} \tag{7}$$

is the resolvent for the complementary space of $P_0$ (namely the intermediate states with higher energies). After reorganizing the electron operators appearing in Equation 6 in terms of spin operators we can find the effective Hamiltonian of the magnetic Raman scattering analogous to Equation 2.

In the perturbative expansion Equation 6, the spins of two neighboring Ni ions (located at **R** and **R′**) can be exchanged with photon absorption/emission. The spins are exchanged along paths via two shared S ions whose positions are designated as follows:

$$\mathbf{R} + \boldsymbol{\delta}_1 = \mathbf{R}' + \boldsymbol{\delta}'_1 = \text{S-ion 1}, \quad \mathbf{R} + \boldsymbol{\delta}_2 = \mathbf{R}' + \boldsymbol{\delta}'_2 = \text{S-ion 2} \tag{8}$$

where $\boldsymbol{\delta}$ is a vector connecting a Ni ion to the surrounding S ion in the hexagon.

The spins can be exchanged in *two distinct mechanisms: via doubly occupied Ni ion or via doubly occupied S ion*. On top of this, the edge-sharing crystal structure allows several spin exchange paths depending on a mechanism. Since the parameters of the tight-binding Hamiltonian are unknown, the relative magnitude/sign of the amplitudes of these two mechanisms are essentially arbitrary. We will further assume that the amplitudes are identical for a given exchange mechanism except for the photon polarization factor. With these provisions, the effective Hamiltonian can be expressed as ($S_R$ is the Ni spin-1 operator at site **R**)

$$H_{eff,NiPS_3} = \sum_{\langle R,R' \rangle} S_R \cdot S_{R'} \left[ M_{Ni} C_{Ni} + M_S C_S \right], \quad (9)$$

where $M_{Ni/S}$ is the amplitude via doubly occupied Ni/S ion *apart from the polarization factor* $C_{Ni/S}$ and $\langle R,R' \rangle$ indicates the nearest neighbor pair. The detailed expressions for $M_{Ni/S}$ are in Supplementary Note 1. The most important polarization factors are as follows ($\hat{\varepsilon}_{i,s}$ is the polarization vector of incoming and scattered photons, respectively):

$$\begin{aligned} C_{Ni} = &\; 2(\hat{\varepsilon}_i \cdot \delta_1)(\hat{\varepsilon}_s^* \cdot \delta_1) + 2(\hat{\varepsilon}_i \cdot \delta_2)(\hat{\varepsilon}_s^* \cdot \delta_2) + 2(\hat{\varepsilon}_i \cdot \delta_1')(\hat{\varepsilon}_s^* \cdot \delta_1') + 2(\hat{\varepsilon}_i \cdot \delta_2')(\hat{\varepsilon}_s^* \cdot \delta_2') \\ &+ 2(\hat{\varepsilon}_i \cdot \delta_1)(\hat{\varepsilon}_s^* \cdot \delta_2) + 2(\hat{\varepsilon}_i \cdot \delta_2)(\hat{\varepsilon}_s^* \cdot \delta_1) + 2(\hat{\varepsilon}_i \cdot \delta_1')(\hat{\varepsilon}_s^* \cdot \delta_2') + 2(\hat{\varepsilon}_i \cdot \delta_2')(\hat{\varepsilon}_s^* \cdot \delta_1'), \end{aligned} \quad (10)$$

$$\begin{aligned} C_S = &\; 2(\hat{\varepsilon}_i \cdot \delta_1)(\hat{\varepsilon}_s^* \cdot \delta_1) + 2(\hat{\varepsilon}_i \cdot \delta_2)(\hat{\varepsilon}_s^* \cdot \delta_2) + 2(\hat{\varepsilon}_i \cdot \delta_1')(\hat{\varepsilon}_s^* \cdot \delta_1') + 2(\hat{\varepsilon}_i \cdot \delta_2')(\hat{\varepsilon}_s^* \cdot \delta_2') \\ &+ 2(\hat{\varepsilon}_i \cdot \delta_1)(\hat{\varepsilon}_s^* \cdot \delta_1) + 2(\hat{\varepsilon}_i \cdot \delta_1)(\hat{\varepsilon}_s^* \cdot \delta_1) + 2(\hat{\varepsilon}_i \cdot \delta_2')(\hat{\varepsilon}_s^* \cdot \delta_2') + 2(\hat{\varepsilon}_i \cdot \delta_2')(\hat{\varepsilon}_s^* \cdot \delta_2'). \end{aligned} \quad (11)$$

In principle, we can expand the spin operator of Equation 9 in a magnon basis starting from the microscopic tight-binding Hamiltonian and use the standard Fermi Golden rule to obtain the scattering intensity. Unfortunately, the magnon operators cannot be obtained within our approximation, but the polarization dependence can still be extracted. An immediate comment is in order: the effective Hamiltonian Equation 9 is *isotropic* in spin, starkly contrasting with the observed planar spin configuration. This is to be expected because the spin-orbit interaction is not considered (the inclusion of spin-orbit interaction renders explicit calculations intractable).

Despite these apparent shortcomings, it turns out that the results for the polarization dependence appear to be consistent with the experimental data with a suitable adjustment of relative weight between $M_{Ni}$ and $M_S$. We point out that the analytical expressions for the exchange constants of the spin Hamiltonian have been obtained in Ref. 36. However, we cannot extract the quantitative values of the exchange constants since the parameters of the multi-orbital tight-binding Hamiltonian are not known.

ASSOCIATED CONTENT

Supporting Information.

The Supporting Information is available free of charge on the ACS Publications website at DOI: 10.1021/acsnano.xxxxxxx.

Calculation of Polarization Dependence; Determination of crystal axes; Low-frequency Raman spectra at 50 K measured by using lasers with different wavelengths; Temperature dependence of polarized Raman spectra for 6L, 5L, 4L, 3L, 2L and 1L $NiPS_3$; Temperature dependence of $M_1$ magnon peak and $\Delta P_2$ for few-layer $NiPS_3$; INS data of $NiPS_3$ along the momentum contour [H, 1/2, 1] (r.l.u.); INS cross-sections of $NiPS_3$ at Q = [0, 1, 0] and [0, 1, 1] calculated from linear spin-wave theory and Equation 4 of the main text.

AUTHOR INFORMATION

Corresponding Authors

*E-mail: hyunlee@sogang.ac.kr, jgpark10@snu.ac.kr, hcheong@sogang.ac.kr

**Author Contributions**

J.-G.P. and H.C. conceived the experiments and supervised the project. J.K. synthesized NiPS$_3$ samples and characterized them for Raman experiments. W.N., S.O., and H.C. prepared few-layer samples and performed Raman measurements. P.P. grew NiPS$_3$ single crystal samples and aligned them for the inelastic neutron scattering measurements, which were conducted with A.S. and A.T. H.C.L carried out the theoretical analysis of Raman scattering. All the authors analyzed the data and wrote the manuscript jointly.

**Notes**

The authors declare no competing financial interest.

ACKNOWLEDGMENT

We acknowledge C. L. Sarkis and M. B. Stone for their help regarding the INS experiment and N. Perkins for helpful discussion. This work was supported by the National Research Foundation (NRF) grant funded by the Korean government (MSIT) (2019R1A2C3006189, 2017R1A5A1014862, SRC program: vdWMRC Center). This research used resources at the Spallation Neutron Source, a U.S. Department of Energy (DOE) Office of Science User Facility operated by the Oak Ridge National Laboratory (ORNL). P. Park acknowledges support by the U.S. Department of Energy, Office of Science, Basic Energy Sciences, Materials Science and Engineering Division. The work at Seoul National University was supported by the Leading Researcher Program of the National Research Foundation of Korea (Grant No. 2020R1A3B2079375). The work by A. Scheie and D.A. Tennant is supported by the Quantum Science Center (QSC), a National Quantum Information Science Research Center of the U.S. Department of Energy (DOE).

For Table of Contents Only

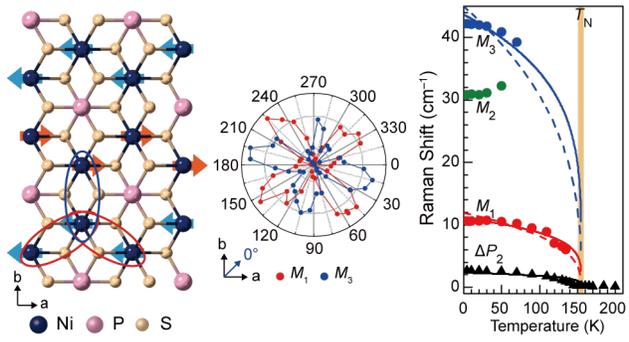

# Supporting Information

# Direct Observation and Analysis of Low-Energy Magnons with Raman Spectroscopy in Atomically Thin NiPS$_3$

*Woongki Na,[†,#] Pyeongjae Park,[‡,§,#] Siwon Oh,[†,#] Junghyun Kim,[‡] Allen Scheie,[⊥] David Alan Tennant,[¶] Hyun Cheol Lee,[†,*] Je-Geun Park[‡,*] Hyeonsik Cheong,[†,*]*

[†]Department of Physics, Sogang University, Seoul 04107, Korea

[‡]Department of Physics and Astronomy, Seoul National University, Seoul 08826, Korea

[§]Materials Science and Technology Division, Oak Ridge National Laboratory, Oak Ridge, Tennessee 37831, USA

[⊥]MPA-Q, Los Alamos National Laboratory, Los Alamos, NM 87545, USA

[¶]Department of Physics and Astronomy, University of Tennessee, Knoxville, TN 37996, USA and Department of Materials Science and Engineering, University of Tennessee, Knoxville, TN 37996, USA

**Contents:**

- **Note S1.** Calculation of Polarization Dependence
- **Figure S1.** Determination of crystal axes.
- **Figure S2.** Low-frequency Raman spectra at 50 K measured by using lasers with different wavelengths.
- **Figure S3.** Temperature dependence of polarized Raman spectra for 6L, 5L, and 4L NiPS$_3$.
- **Figure S4.** Temperature dependence of polarized Raman spectra for 3L, 2L, and 1L NiPS$_3$.
- **Figure S5.** Temperature dependence of M$_1$ magnon peak and ΔP$_2$ for few-layer NiPS$_3$.
- **Figure S6.** INS data of NiPS$_3$ along the momentum contour [H, 1/2, 1] (r.l.u.).
- **Figure S7.** INS cross-sections of NiPS$_3$ at Q = [0, 1, 0] and [0, 1, 1] calculated from linear spin-wave theory and Equation 4 of the main text.

## Note S1. Calculation of Polarization Dependence

### 1) Parametrization of lattice sites

The two-dimensional crystal structure[1] of NiPS$_3$ can be parameterized as follows. The Ni ions form a honeycomb lattice, which can be described as a triangular lattice with a basis. Let $a$ be the distance between Ni ions, then the triangular lattice can be spanned by

$$\vec{a}_1 = \sqrt{3}a\hat{\mathbf{a}}, \quad \vec{a}_2 = \frac{\sqrt{3}}{2}a\hat{\mathbf{a}} + \frac{3}{2}a\mathbf{b}, \tag{S1}$$

$\hat{\mathbf{a}}$ is the unit vector along **a**-axis, etc. The honeycomb lattice can be described with a basis at $\eta = a\hat{\mathbf{b}}$.

Honeycomb lattice: $R_{m_1,m_2} = m_1\vec{a}_1 + m_2\vec{a}_2, \quad R_{m_1,m_2} + \eta.$

Second, put a Ni ion at the center of a hexagon formed by 6 surrounding S ions. This implies that the trigonal axis of the undeformed octahedron is along the $c$-axis. The distance in the $ab$-plane between Ni ion and S ion is $a/\sqrt{3} = a'$. The planar positions of the six S ions relative to the center, which is taken to be the origin, can be written as follows:

$$\begin{aligned}
\delta_{1+} &= a'\hat{\mathbf{a}}, \quad \delta_{2-} = \frac{a'}{2}\hat{\mathbf{a}} + \frac{\sqrt{3}}{2}a'\mathbf{b}, \\
\delta_{3+} &= -\frac{a'}{2}\hat{\mathbf{a}} + \frac{\sqrt{3}}{2}a'\mathbf{b}, \quad \delta_{4-} = -a'\hat{\mathbf{a}}, \\
\delta_{5+} &= -\frac{a'}{2}\hat{\mathbf{a}} - \frac{\sqrt{3}}{2}a'\mathbf{b}, \quad \delta_{6-} = +\frac{a'}{2}\hat{\mathbf{a}} - \frac{\sqrt{3}}{2}a'\mathbf{b},
\end{aligned} \tag{S2}$$

where $\pm$ subscripts indicate whether the S-ion is above or below the Ni plane.

We note that this crystal structure is edge-sharing, and so each S ion can be connected to two Ni ions by nearest-neighbor hopping. Now, the S-ion sites can be parametrized by

$$R_{m_1,m_2,(0,\eta)} + \delta_{i\pm}. \tag{S3}$$

### 2) Tight-binding model Hamiltonian

We transcribe the Hamiltonian by Koshibae *et al.*[2], which is being adapted to the crystal structure of

NiPS$_3$. The lattice sites have been fully parameterized in Section 1.1, and they will be denoted compactly in this section to avoid cumbersome notations. Let **R** and **R**$p$ denote a generic position of Ni and S ions, respectively. The tight-binding Hamiltonian consists of the on-site and hopping terms (the electron-photon interaction Hamiltonian will be treated in the next section).

$$\hat{H}_{\text{t-b}} = \hat{H}_{\text{on-site}} + \hat{H}_{\text{hop}} \tag{S4}$$

Below $d^\dagger_{\mathbf{R}\alpha\sigma}$ and $p^\dagger_{\mathbf{R}_p a\sigma}$ are the *hole creation operators* with spin $\sigma$ for Ni ion and S ion at site **R**, **R**$_p$, respectively. $\alpha = 0, 1$ is the $e_g$ orbital index, and $a = p_x, p_y, p_z$ is the p-orbital index. The number operators are defined as

$$n^d_{\mathbf{R}\alpha\sigma} = d^\dagger_{\mathbf{R}\alpha\sigma} d_{\mathbf{R}\alpha\sigma}, \quad n^p_{\mathbf{R}_p a\sigma} = p^\dagger_{\mathbf{R}_p a\sigma} p_{\mathbf{R}_p a\sigma}. \tag{S5}$$

The spin summed number operator is denoted by $n^d_{\mathbf{R}\alpha}$, etc. The on-site Hamiltonian consists of the site-energy term, Coulomb repulsion term, and spin-orbit interaction term.

$$\begin{aligned}
\hat{H}_{\text{on-site}} &= \sum_{\mathbf{R},\alpha,\sigma} \epsilon_\alpha n^d_{\mathbf{R}\alpha\sigma} + \sum_{\mathbf{R}_p,a,\sigma} \epsilon_a n^p_{\mathbf{R}_p a\sigma} & \text{site energy} \\
&+ U \sum_{\mathbf{R},\alpha} n^d_{\mathbf{R}\alpha\uparrow} n^d_{\mathbf{R}\alpha\downarrow} & \text{Hubbard Coulomb repulsion} \\
&+ U_{01} \sum_{\mathbf{R}} n^d_{\mathbf{R}0} n^d_{\mathbf{R}1} & \text{inter-orbital Coulomb repulsion} \\
&+ \lambda \sum_{\mathbf{R}} \left( \mathbf{L}^{\text{I}}_{\mathbf{R}} \cdot \mathbf{S}^{\text{I}}_{\mathbf{R}} + \mathbf{L}^{\text{II}}_{\mathbf{R}} \cdot \mathbf{S}^{\text{II}}_{\mathbf{R}} \right) & \text{spin-orbit interaction,}
\end{aligned} \tag{S6}$$

where I and II denote two holes on a Ni ion, and other notations are self-explanatory. $\epsilon_0 \approx \epsilon_1$ is assumed for d-orbitals, while the p-orbital site energies can be split by trigonal distortion. All parameters of the Hamiltonian should be regarded as phenomenological. The explicit form of the operators of spin-orbit interaction is given by

$$\mathbf{L}_{\mathbf{R}} \cdot \mathbf{S}_{\mathbf{R}} = \sum_{i=x,y,z} (L^i)_{\alpha\alpha'} \left(\frac{\sigma^i}{2}\right)_{\sigma\sigma'} d^\dagger_{\mathbf{R}\alpha\sigma} d_{\mathbf{R}\alpha'\sigma'}. \tag{S7}$$

$(L^i)_{\alpha\alpha'}$ is the matrix element of orbital angular momentum $l = 2$ operator, and $\sigma^i$ is the i-th Pauli sigma matrix.

The spin-orbit coupling is treated perturbatively. In the first order, it can modify the hopping amplitude to become spin-dependent. This contribution will induce a spin-dependent electron-photon coupling, which we ignore for simplicity (analytical computations turn out to be too formidable). In the second order, it gives rise to the *single ion anisotropy*, which is responsible for the planar spin configuration[3], while it does not couple to photon because it is an on-site interaction.

The hopping Hamiltonian between Ni and S ion is given by

$$\hat{H}_{hop} = \sum_{<\mathbf{R}, \mathbf{R}_p>a,\alpha,\sigma} \left( t^{\alpha a}_{\mathbf{R},\mathbf{R}_p} d^\dagger_{\mathbf{R}\alpha\sigma} p_{\mathbf{R}_p,a,\sigma} + \text{h.c.} \right), \tag{S8}$$

where $t^{\alpha a}_{\mathbf{R},\mathbf{R}_p}$ is the nearest neighbor hopping amplitude and h.c. stands for the Hermitian conjugate.

### 3) Electron-photon interaction Hamiltonian

In visible range, the position dependence of the vector potential can often be neglected so that the line integral in the exponent of the Peierls phase can be simplified to

$$\exp\left(-i\frac{e}{\hbar c}(\mathbf{R}_i - \mathbf{R}_j)\cdot \mathbf{A}(t)\right) \tag{S9}$$

Then the Peierls coupling to the hopping Hamiltonian Equation S8 takes the following form:

$$\hat{H}_{hop+em} = \sum_{<\mathbf{R}, \mathbf{R}_p>a,\alpha,\sigma} \left( t^{\alpha a}_{\mathbf{R},\mathbf{R}_p} \exp\left[-i\frac{e}{\hbar c}\mathbf{A}(t)\cdot(\mathbf{R}-\mathbf{R}_p)\right] d^\dagger_{\mathbf{R}\alpha\sigma} p_{\mathbf{R}_p,a,\sigma} + \text{h.c.} \right). \tag{S10}$$

The expansion of $H_{hop+em}$ with respect to the vector potential $\mathbf{A}(t)$ in the first order gives the electron-photon interaction Hamiltonian for the magnetic Raman scattering.

$$\hat{H}_{e\text{-photon}} = -\frac{e}{c}\mathbf{A}(t)\cdot \mathbf{j}, \tag{S11}$$

where the current operator is

$$\mathbf{j} = \frac{i}{\hbar}\sum_{<\mathbf{R}, \mathbf{R}_p>a,\alpha,\sigma} (\mathbf{R}-\mathbf{R}_p) t^{\alpha a}_{\mathbf{R},\mathbf{R}_p} (d^\dagger_{\mathbf{R}\alpha\sigma} p_{\mathbf{R}_p,a,\sigma} - \text{h.c.}). \tag{S12}$$

The direct coupling between the photon polarization and the crystal structure manifests in Equation S11. This implies that the deformation of the crystal structure is directly reflected in the polarization dependence. We again emphasize that the spin-orbit interaction is not considered in our theoretical calculation of the effective scattering Hamiltonian.

## 4) Some details of perturbation calculations

A key identity in the derivation of the effective scattering Hamiltonian is

$$\sum_{\sigma,\sigma'} d^\dagger_{\mathbf{R}\alpha\sigma'} d_{\mathbf{R}'\alpha'\sigma'} d^\dagger_{\mathbf{R}'\alpha'\sigma} d_{\mathbf{R}\alpha\sigma} = (-2)\mathbf{S}_{\mathbf{R}\alpha} \cdot \mathbf{S}_{\mathbf{R}'\alpha'} + \text{constant}, \tag{S13}$$

which is valid in the low-energy Hilbert space. By taking the symmetric combination of the above operator and using the Wigner-Eckart theorem[2] we can obtain the effective Hamiltonian in terms of spin-1 operator.

Let us consider a scattering process in which a photon is absorbed first. In this case, some of the explicit expressions of resolvents Equation 7 of the main text are as follows ($\epsilon_d = \epsilon_{0,1}$, and $\hbar\omega_i$ is the energy of the incident photon ).

$$G = \frac{1}{\epsilon_d + U_{01} + \hbar\omega_i - \epsilon_{p_a}}, \quad \text{one S-ion hole with orbital } p_a \text{ being occupied.} \tag{S14}$$

$$G = \frac{1}{\hbar\omega_i - U}, \text{ one orbital of Ni ion being doubly occupied.} \tag{S15}$$

$$G = \frac{1}{2\epsilon_d + 2U_{01} + \hbar\omega_i - \epsilon_{p_a} - \epsilon_{p_b}}. \text{ Two S ion holes in orbitals } p_a, p_b \text{ being doubly occupied} \tag{S16}$$

The case where the photon emission occurs first can be handled similarly. Using the above results, we can compute the amplitudes $\mathcal{M}_{\text{Ni/S}}$ of Equation 9 of the main text.

One example of the exchange amplitude via doubly occupied Ni ion with the following exchange path $\mathbf{R} \to \mathbf{R}+\boldsymbol{\delta}_1 \to \mathbf{R}' \to \mathbf{R}+\boldsymbol{\delta}_1 \to \mathbf{R}$ is given by

$$\sum_{a,b} \frac{1}{\epsilon_d + U_{01} + \hbar\omega_i - \epsilon_{p_a}} \frac{1}{\hbar\omega_i - U} \frac{1}{\epsilon_d + U_{01} + \hbar\omega_i - \epsilon_{p_b}} t_{0a|\delta_1} t_{0a|\delta_1'}, t_{0b|\delta_1'} t_{0b\delta_1} [\mathbf{A}_i \cdot \boldsymbol{\delta}_1][\mathbf{A}_s^* \cdot (-\boldsymbol{\delta}_1)], \tag{S17}$$

where $\mathbf{A}_{i,s}$ is the vector potential of the incoming and the scattered photon, respectively.

Next, let us consider the exchange amplitude via doubly occupied S ion with the following

exchange path

$$\mathbf{R} \to \mathbf{R} + \boldsymbol{\delta}_1 \to \mathbf{R} \quad \text{and} \quad \mathbf{R}' \to \mathbf{R}' + \boldsymbol{\delta}_{1'} \to \mathbf{R}' \tag{S18}$$

In the above path, the S-ion 1 is doubly occupied, and the photon is absorbed and emitted in the hop $\mathbf{R} \leftrightarrow \mathbf{R} + \boldsymbol{\delta}_1$. The associated amplitude is given by

$$\sum_{a,b} \frac{1}{\epsilon_d + U_{01} + \hbar\omega_i - \epsilon_{p_a}} \frac{1}{2\epsilon_d + 2U_{01} + \hbar\omega_i - \epsilon_{p_a} - \epsilon_{p_b}} \left( \frac{1}{\epsilon_d + U_{01} + \hbar\omega_i - \epsilon_{p_a}} + \frac{1}{\epsilon_d + U_{01} + \hbar\omega_i - \epsilon_{p_b}} \right) \tag{S19}$$
$$\times t_{0a}^2 t_{0b}^2 [\mathbf{A}_i \cdot \boldsymbol{\delta}_1][\mathbf{A}_s^* \cdot (-\boldsymbol{\delta}_1)]$$

There are many other possible exchange paths, and the corresponding amplitudes can be computed in an entirely identical way.

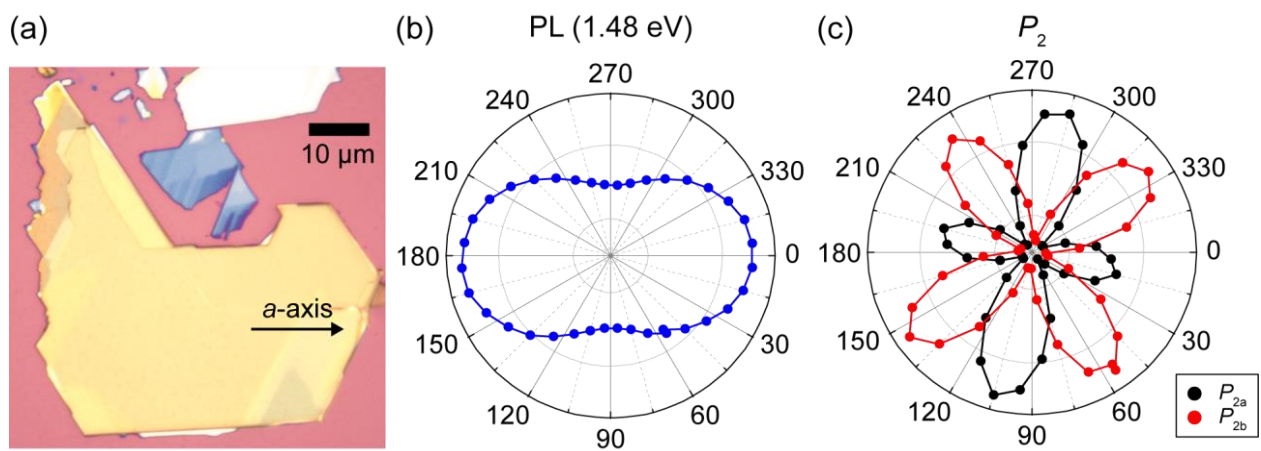

**Figure S1.** Determination of crystal axes. a) Optical microscope image of sample. b,c) Polarization dependence of the PL at 1.48 eV and the split $P_2$ peaks at 3.5 K.

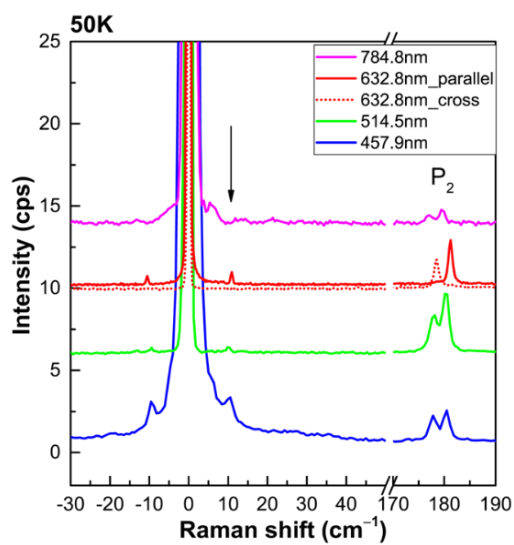

**Figure S2.** Low-frequency Raman spectra at 50 K measured by using lasers with different wavelengths.

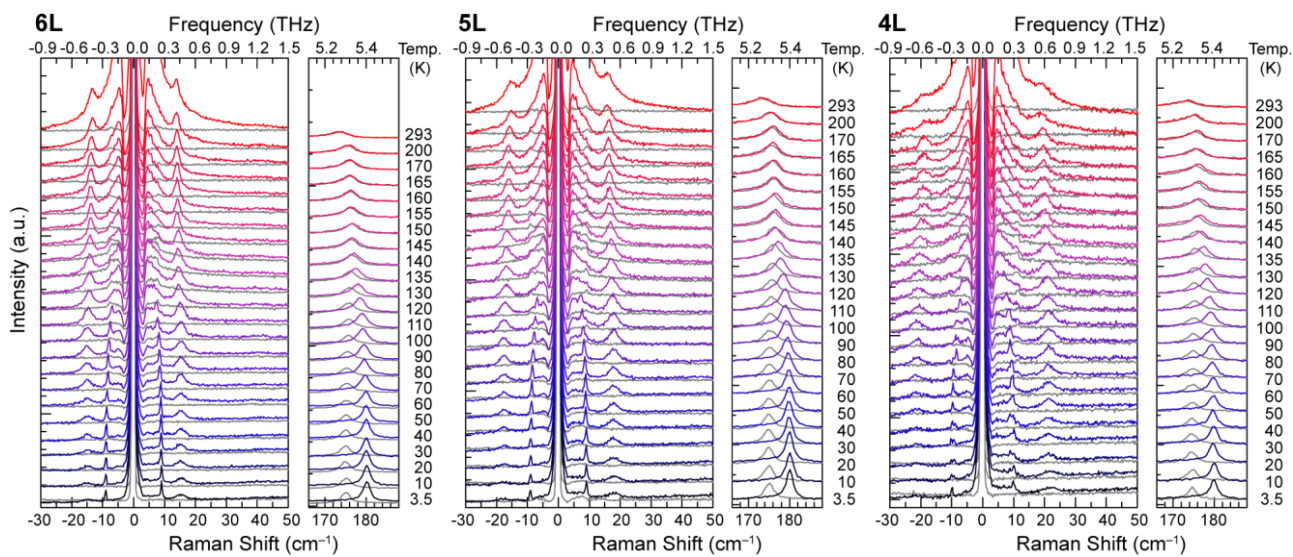

**Figure S3.** Temperature dependence of polarized Raman spectra for 6L, 5L, and 4L NiPS$_3$. The colored spectra are measured in the $z(xx)\bar{z}$ polarization configuration and the grey spectra in $z(xy)\bar{z}$.

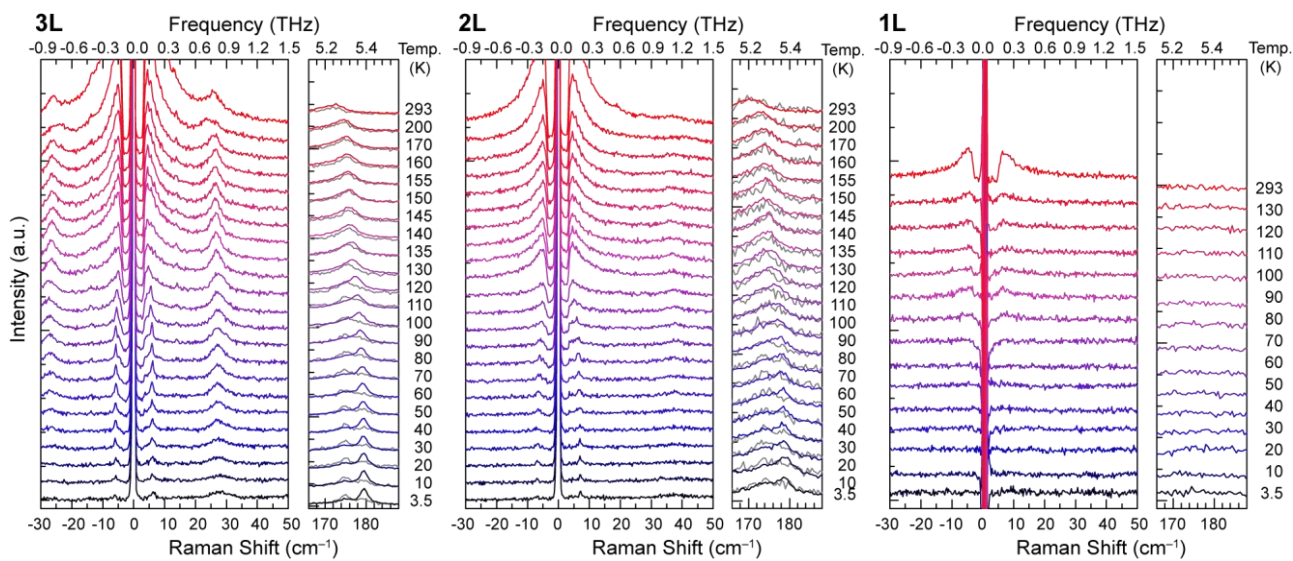

**Figure S4.** Temperature dependence of polarized Raman spectra for 3L, 2L, and 1L NiPS$_3$ measured in the $z(xx)\bar{z}$ polarization configuration.

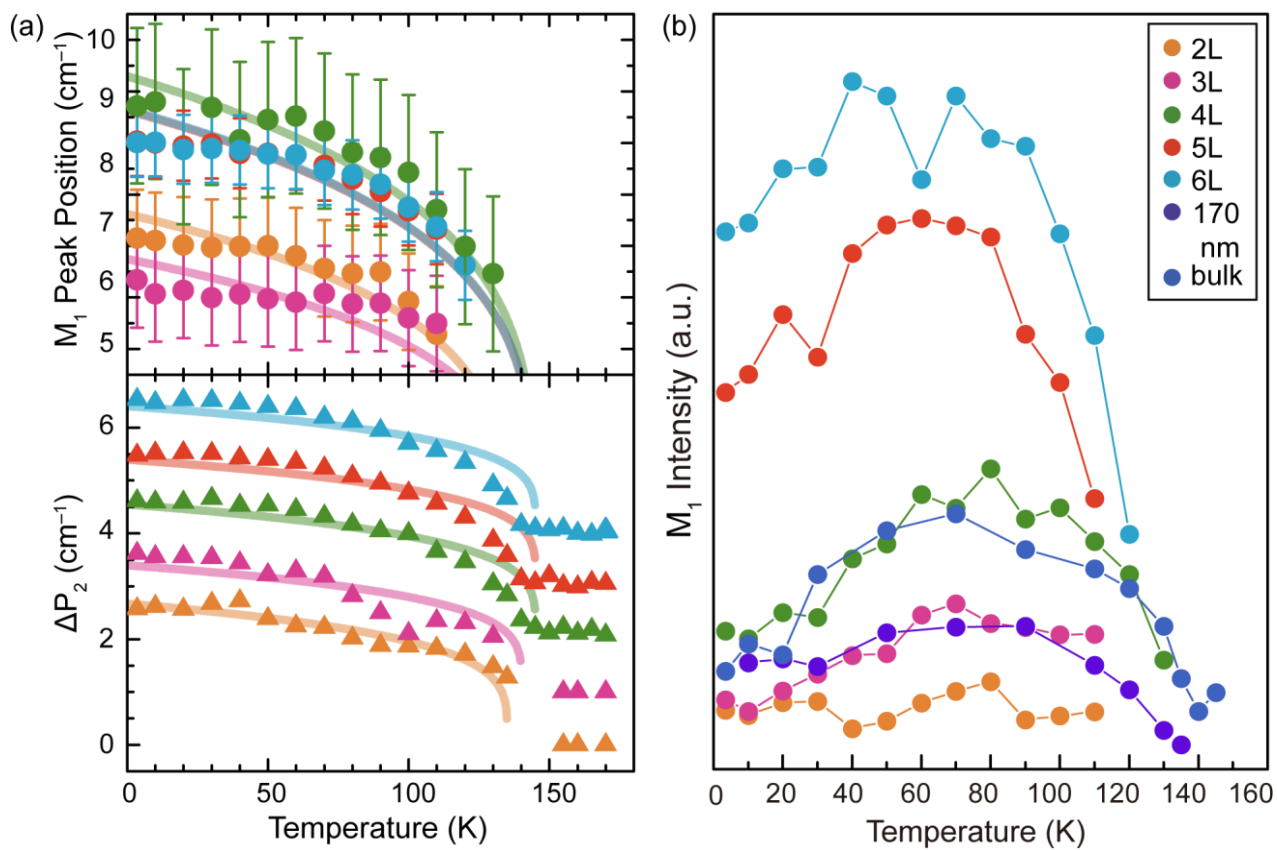

**Figure S5.** a) Temperature dependence of $M_1$ magnon peak position and $\Delta P_2$ for few-layer $NiPS_3$. The data for $\Delta P_2$ are offset vertically for clarity. The curves show temperature dependence of $[1-T/T_N]^\beta$ with the exponent of 0.23. b) Temperature dependence of $M_1$ magnon peak intensity.

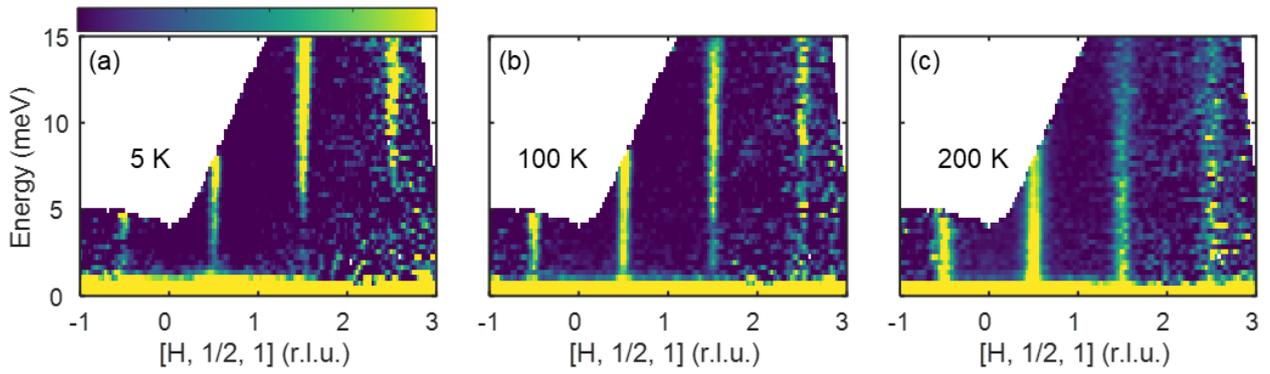

**Figure S6.** INS data of NiPS$_3$ along the momentum contour [H, 1/2, 1] (r.l.u.). Magnon dispersion down to ~ 1.5 meV is clearly visible at [1/2, 1/2, 1].

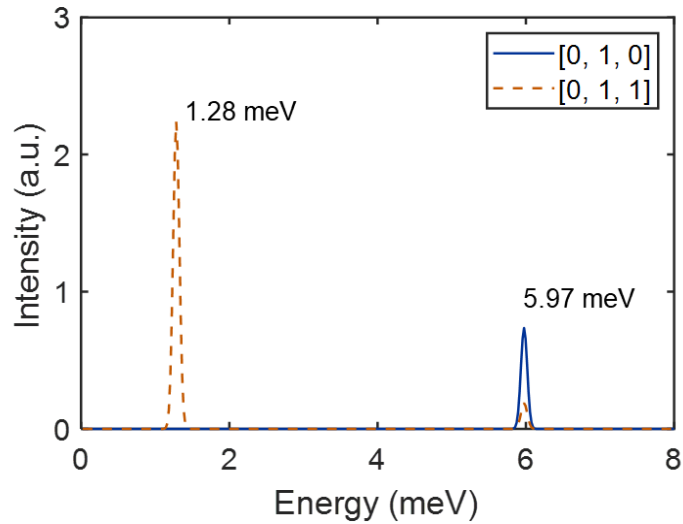

**Figure S7.** INS cross-sections of NiPS$_3$ at Q = [0, 1, 0] (solid blue curve) and [0, 1, 1] (dashed orange curve) calculated from linear spin-wave theory and Equation 4 of the main text. For a clear demonstration of magnon eigenvalues, the cross-sections were convoluted by a Gaussian function with the full-width-at-half-maximum of 0.1 meV. Note that the calculation results shown in Figure 5c-f are different from those shown in this figure as they include the effects of the momentum resolution and the experimental bin width.